\documentclass[prl,aps,twocolumn,footinbib,superscriptaddress,floatfix,bibliography]{revtex4-1}

\usepackage{CJK}
\usepackage{color} 
\usepackage{graphicx}
\usepackage{float}
\usepackage{multirow}
\usepackage{amsmath}
\usepackage{bm}
\usepackage{amsfonts}
\usepackage{braket}
	
\usepackage[caption=false,subrefformat=parens,labelformat=parens]{subfig}

\usepackage[dvipsnames]{xcolor}
\usepackage{hyperref}
\usepackage{cleveref}

\newcommand\myshade{85}
\colorlet{mylinkcolor}{violet}
\colorlet{mycitecolor}{YellowOrange}
\colorlet{myurlcolor}{Aquamarine}

\hypersetup{
	linkcolor  = myurlcolor!\myshade!black,
	citecolor  = mycitecolor!\myshade!black,
	urlcolor   =  mylinkcolor!\myshade!black,
	colorlinks = true,
}

\DeclareMathOperator{\Tr}{Tr}

\newcommand{\bfm}{{\bf m}}
\newcommand{\bfq}{{\bf q}}

\newcommand{\bfs}{{\bf S}}

\newcommand{\BYZO}{Ba$_3$Yb$_2$Zn$_5$O$_{11}$}

\makeindex
\begin{document}
\graphicspath{{./newfigs/}}
\begin{CJK*}{UTF8}{gbsn} 

\title{Rank--2 $U(1)$ spin liquid on the breathing pyrochlore lattice}

\author{Han Yan (闫寒)}
\email{han.yan@oist.jp}
\affiliation{Okinawa Institute of Science and Technology Graduate University, Onna-son, 
	Okinawa 904-0412, Japan}
\author{Owen Benton}
\affiliation{RIKEN Center for Emergent Matter Science (CEMS), Wako, Saitama, 351-0198, Japan}
\author{Ludovic D.C. Jaubert}
\affiliation{CNRS, Universit\'e de Bordeaux, LOMA, UMR 5798, 33400 Talence, France}
\author{Nic Shannon}
\affiliation{Okinawa Institute of Science and Technology Graduate University, Onna-son, 
Okinawa 904-0412, Japan}

\date{\today}

\begin{abstract}

Higher--rank generalisations of electrodynamics have recently attracted considerable
attention because of their ability to host ``fracton'' excitations, with connections to 
both fracton topological order and 
gravity.  
However, the search for higher--rank gauge theories in experiment has 
been greatly hindered by the lack of materially--relevant microscopic models.
Here we show how a spin liquid described by rank--2 $U(1)$ gauge theory can
arise in a magnet on the breathing pyrochlore lattice.
We identify Yb--based breathing pyrochlores as candidate systems, 
and make explicit predictions for how the rank--2 $U(1)$ spin liquid 
would manifest itself in experiment.

\end{abstract}
\maketitle
\end{CJK*}


{\it Introduction. } 
It is of great intellectual interest and practical utility to discover novel effective laws 
of nature emerging from many--body systems.
Traditionally, this enterprise has been entwined with the concept of broken 
symmetry \cite{anderson72}. 
However, a powerful alternative has proved to be the local constraints 
which arise from competing or ``frustrated'', interactions.
In the context of frustrated magnets, these can lead to the emergence of 
a local gauge symmetry, and thereby to quantum spin liquids, which defy 
all usual concepts of magnetic order, and instead 
exhibit fractionalised excitations and long--range entanglement 
\cite{Anderson1973,Balents2010,Savary2016,Zhou2017}.
A well--studied 
example is quantum spin ice, 
a realisation of a $U(1)$ gauge theory on the pyrochlore lattice, 
whose emergent excitations exactly mimic conventional 
electrodynamics: photons, electric charges and magnetic monopoles.
As such, it
has attracted intense theoretical  \cite{hermele04PRB,banerjee08-PRL100,benton12-PRB86,savary12,
	Shannon2012PRL,hao14,Gingras14RoPP,kato15,chen17,huang18-PRL120} 
and experimental 
\cite{zhou08,RossPRX11,fennell12,Kimura2013,sibille15,wen17,thompson17,Sibille2018,gao-arXiv} investigation.


Recent work has highlighted the possibility of more exotic forms 
of emergent electrodynamics \cite{XuPRB06,PretkoPRB16,Rasmussen2016arXiv,PretkoPRB17}, 
where electric and magnetic fields have the form of rank--2 (or higher--rank) tensors.
These theories have modified conservation laws and gauge symmetries, resulting in
some remarkable properties.
Some are argued to mimic gravity \cite{XuPRB06,Benton2016NatComm,PretkoPRD16},
while others are dual to elasticity theory \cite{PretkoPRL18, gromov19}.
In both cases, the charged excitations, dubbed ``fractons'', have constrained mobility, 
and characterize a new class of topological order  \cite{chamon05-PRL94,shannon04-PRB69,HaahPRA11,VijayPRB15,VijayPRB2016,BulmashPRB2018,MaPRB2018,Nandkishore2018arXiv,SlaglePRB17,GaborPRB17}.
Fracton models are also linked to quantum stabilizer codes \cite{SchmitzPRB2018,KubicaArXiv18}
and holography \cite{Yan2018arXiv}. 
None the less, these desirable properties come at a price: the local constraint required 
has a tensor character.
As a consequence, prototypical models of fractons  require rather complicated interactions \cite{chamon05-PRL94,XuPRB06,HaahPRA11,VijayPRB15,VijayPRB2016}, 
with just a handful of proposals motivated by experiment  \cite{SlaglePRB17,GaborPRL17,You2018arXiv}.
In the case of gapless higher--rank gauge theories, only a few concrete models exist \cite{XuPhysRevD2010,XuPhysRevB2007,Rasmussen2016arXiv}, 
and even less is known about how to achieve such a phase in a real material.
For this reason, realizing an emergent higher--rank electrodynamics in 
experiment presents a significant challenge.


\begin{figure}[t]
	\centering
	\subfloat[BP Lattice\label{fig:lattice}]{
		\includegraphics[height=0.3\columnwidth]{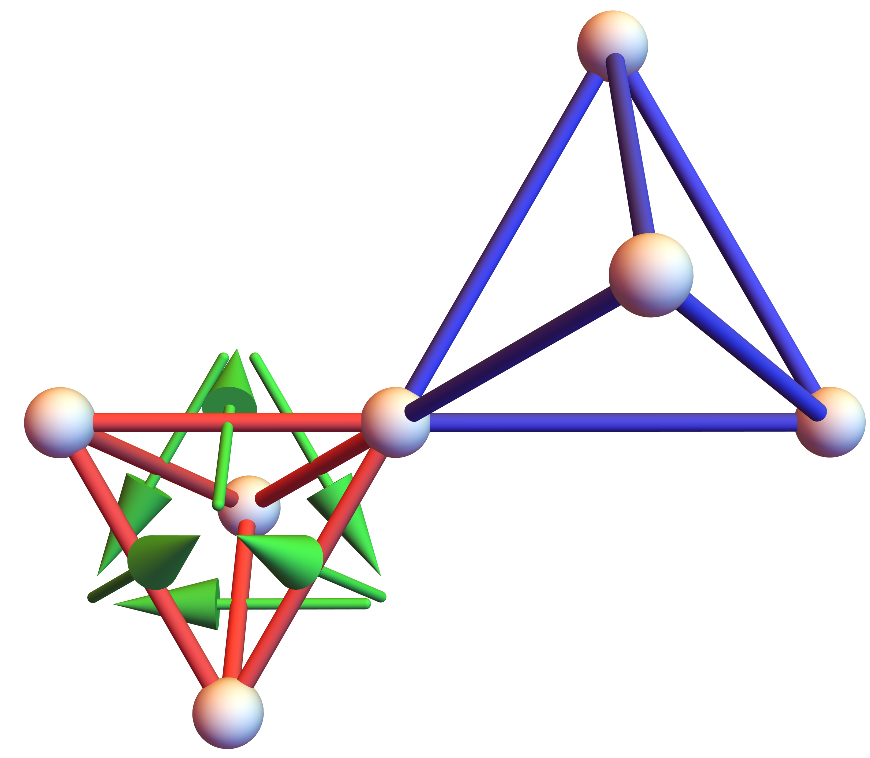}} \;
	\subfloat[0kl plane \label{fig:2FPP}]{
		\includegraphics[height=0.32\columnwidth]{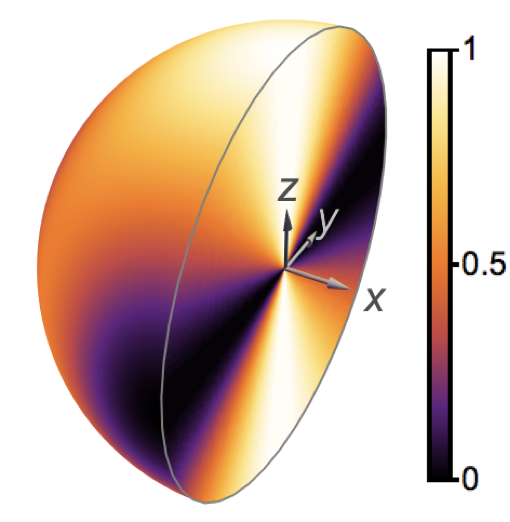}} \\
	\subfloat[hk0 plane \label{fig:4FPP}]{
		\includegraphics[width=0.35\columnwidth]{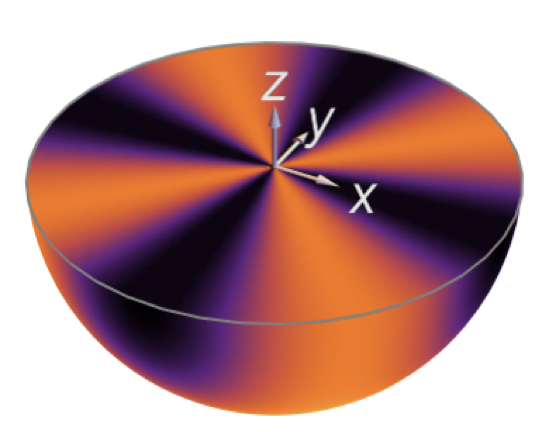}} \;
	\subfloat[$\langle E_{\sf xy}({\bf q}) E_{\sf xy}(-{\bf q}) \rangle$ \label{fig:4FPP-MC}]{
		\includegraphics[height=0.3\columnwidth]{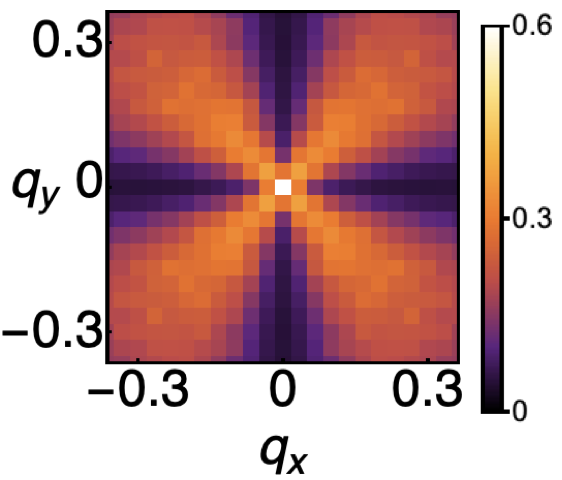}} 
	%
	%
	\caption{
		Breathing pyrochlore (BP) lattice, and singular correlations
		characteristic of a rank--2 $U(1)$ [\mbox{R2--U1}] gauge theory.
		%
		(a) 
		BP lattice, with A-- and B--sublattice tetrahedra of unequal size.  
		The vectors associated with Dzyaloshinskii--Moriya (DM) interactions on the 
		A--sublattice [Eq.~(\ref{eq:H})] are illustrated with green arrows.
		(b) Prediction of \mbox{R2--U1} theory for the correlation function 
		$\langle E_{\sf xy} ({\bf q}) E_{\sf xy} ({\bf -q}) \rangle$ [Eq.~(\ref{eq:4FPP})], 
		showing a 2--fold pinch point in the [0kl] plane.
		(c) Perpendicular section, showing a 4--fold pinch point (4FPP) in the [hk0] plane.
		(d) Equivalent results from MC simulation of the 
		breathing pyrochlore model [Eq.~(\ref{eq:H})]. 
	}
	\label{fig:1}
\end{figure}


In this Letter we show how a canonical rank--2 $U(1)$ [\mbox{R2--U1}] 
spin liquid can arise in a realistic model of a frustrated magnet.
The model we consider is the Heisenberg antiferromagnet (HAF) on a 
breathing--pyrochlore (BP) lattice, perturbed by weak Dzyaloshinskii--Moriya (DM) 
interactions [Fig.~\ref{fig:lattice}].
Working in the classical limit, relevant to a spin liquid at finite temperatures, 
we establish that  
fluctuations can be described using a tensor field satisfying 
the constraints required for a \mbox{R2--U1} gauge theory.
We use classical Monte Carlo (MC) simulation to confirm this scenario, 
and to explore how a \mbox{R2--U1} spin liquid could be identified in experiment.
We find that 4--fold pinch points (4FPP), characteristic of the \mbox{R2--U1} 
state \cite{PremPRB18}, become visible in polarised neutron scattering.
We discuss the application of these ideas to real materials,
identifying Yb--based breathing pyrochlores as potential candidates 
for an \mbox{R2--U1} spin liquid state.
These results complement earlier work exploring gapped, fracton 
topological order, in models with bilinear interactions 
\cite{SlaglePRB17,GaborPRL17,You2018arXiv}, providing an example of
an \mbox{R2--U1} state, in an experimentally--motivated context. 


{\it Review of \mbox{R2--U1} theory. }
Conventional, $U(1)$, electrodynamics  is built around a 
vector field $\bf E$, subject to a Gauss law 
$
\partial_i E_i = \rho \; 
$
so that, in the absence of charges, 
\begin{equation}
\partial_i E_i = 0 \; .
\label{eq:spin.ice.constraint}
\end{equation}
The key which unlocked the effective electrodynamics of spin ice was the realisation 
that, at low temperatures, in a classical limit, spins satisfied a local constraint 
of precisely 
the form of Eq.~(\ref{eq:spin.ice.constraint}) \cite{harris97,bramwell01}.


Here we consider instead an \mbox{R2--U1} electrodynamics, in its self--dual, 
vector--charged, traceless form \cite{PretkoPRB16,Rasmussen2016arXiv,PretkoPRB17}, 
and seek to show that, in an equivalent classical limit, spins satisfy the appropriate generalisation of 
Eq.~(\ref{eq:spin.ice.constraint}).
The \mbox{R2--U1} theory is built around a rank--2 tensor electric field $\bf E$ 
that is symmetric and traceless,
\begin{equation} 
E_{ji}  = E_{ij} \quad \Tr {\bf E} = 0 \; ,
\label{eq:constraint.1}
\end{equation}
subject to a generalised Gauss' law for a vector charge 
\begin{equation}
\partial_i E_{ij} = \rho_j \; .
\end{equation}
In the low--energy sector, the theory is charge free, i.e.
\begin{equation}
\partial_i E_{ij} = 0 \; .
\label{eq:constraint.2}
\end{equation}
These constraints determine the symmetry of the \mbox{R2--U1} gauge field
\begin{equation}
\label{EQN_gauge_symmetry}
A_{ij} \rightarrow A_{ij} + \partial_i \lambda_j + \partial_j \lambda_i + \gamma \delta_{ij} \;,
\end{equation}
which in turn implies the form of the associated magnetic field,  $B_{ij}$ \cite{PretkoPRB16,PretkoPRB17}.
However the key observable properties of an \mbox{R2--U1} spin liquid follow 
from the correlations of its electric field $E_{ij}$ \cite{PremPRB18}, 
and our goal will therefore be to show how the spins in a frustrated magnet can be 
described by a tensor field  $E_{ij}$, satisfying the constraints 
Eqs.~(\ref{eq:constraint.1}, \ref{eq:constraint.2}).


{\it The model. } 
To this end, we consider a HAF, perturbed by weak DM interactions, 
on a ``breathing'' pyrochlore (BP) lattice, for which A-- and B--sublattice tetrahedra 
have a different size
\begin{equation}
\begin{split}
\mathcal{H}_{\sf BP} =& 
\sum_{\langle ij \rangle\in \text{A}} \left[J_A \bfs_i \cdot \bfs_j 
+
D_A\hat{\bf d}_{ij}\cdot( \bfs_i \times \bfs_j )\right ] \\
& + 
\sum_{\langle ij \rangle\in \text{B}} \left[J_B \bfs_i \cdot \bfs_j 
+
D_B  \hat{\bf d}_{ij}\cdot( \bfs_i \times \bfs_j )\right ] \; .
\end{split}
\label{eq:H}
\end{equation}
Definitions of the bond--dependent vectors $\hat{\bf d}_{ij}$ \cite{KotovPRB05,PooleJPCM07,ElhajalPRB05,CanalsPRB08}
are given in the Supplemental Material \mbox{[cf. Fig.~\ref{fig:lattice}]}.
This model  finds experimental motivation in Yb--based 
breathing pyrochlores, discussed below.
%


{\it Transcription to symmetry--based coordinates. } 
Our next step is to seek a continuum representation of Eq.~\eqref{eq:H}.
To accomplish this, we consider the classical limit where individual 
components of spin commute, and introduce a set of coarse--grained fields 
${\bf m}_{\mathsf{X}}$ which transform as irreducible representations 
of the lattice symmetry \cite{BentonThesis,YanPRB17,Benton2016NatComm}.
In this basis \footnote{See Supplemental Materials for a more detailed derivation.}, 
the Hamiltonian becomes 
\begin{equation}
\mathcal{H} 
= \frac{1}{2} \sum_{{\sf tet \in A}, \mathsf{X}} 
a_{\text{A}, \mathsf{X}} m^2_{\mathsf{X}} 
+ \frac{1}{2} \sum_{{\sf tet \in B}, \mathsf{X}}  
a_{\text{B}, \mathsf{X}} m^2_{\mathsf{X}} \; ,
\label{eq:H.m.Td}
\end{equation}
where $\mathsf{X}$ runs over irreps of the group $T_d$, i.e. $\{ \mathsf{A_2}, \mathsf{E}, \mathsf{T_2}, \mathsf{T_{1+}}, \mathsf{T_{1-}} \}$, with the fields $m_\mathsf{X}$ 
and the coefficients $a_{ \sf X}$ defined in Table~I and Table~II of the Supplementary Material.


Before considering the effect of DM interactions, 
it is helpful to explore how this approach works in the case of a known spin liquid, 
the HAF on a pyrochlore lattice \cite{anderson56,reimers91,moessner98-PRL80,moessner98-PRB58,henley05,henley10}.
%
%
Setting 
\begin{eqnarray}
J_A=J_B \; , \; D_A=D_B=0 \; , 
\end{eqnarray}
we find 
\begin{eqnarray}
0 <  a_\mathsf{A_2} = a_\mathsf{E} = a_\mathsf{T_2} = a_\mathsf{T_{1-}} < a_\mathsf{T_{1+}}  \; .
\label{eq:HAF.condition}
\end{eqnarray}
%
%
It follows that the fields $m_\mathsf{A_2}, m_\mathsf{E}, m_\mathsf{T_2}, m_\mathsf{T_{1-}}$ 
are all free to fluctuate in the ground state.  
We can conveniently collect all of these fields 
in the rank--2 tensor
\begin{equation}
{\bf E}^{\sf HAF} 
=  {\bf E}^{\sf HAF}_{\sf sym.} 
+  {\bf E}^{\sf HAF}_ {\sf antisym.}
+ {\bf E}^{\sf HAF}_{\sf trace}
\end{equation}
where 
\begin{equation}
{\bf E}^{\sf HAF}_{\sf sym.}  = 
\begin{bmatrix}
\frac{2}{\sqrt{3}}m_\mathsf{E}^1  &  m_\mathsf{T_{1-}}^z &  m_\mathsf{T_{1-}}^y  \\
m_\mathsf{T_{1-}}^z  & -\frac{1}{\sqrt{3}}m_\mathsf{E}^1 - m_\mathsf{E}^2  &   m_\mathsf{T_{1-}}^x  \\
m_\mathsf{T_{1-}}^y &  m_\mathsf{T_{1-}}^x   &  -\frac{1}{\sqrt{3}}m_\mathsf{E}^1 + m_\mathsf{E}^2
\end{bmatrix}   ,
\label{eq:Ematrix}
\end{equation}
\begin{equation}
( E^{\sf HAF}_ {\sf antisym.})_{ij} = - \epsilon_{ijk} m_\mathsf{T_2}^k,
\quad 
({E}^{\sf HAF}_{\sf trace})_{ij}  = -\delta_{ij}\sqrt{\frac{2}{3}}m_\mathsf{A_2} .
\end{equation}
The requirement of the continuity of the fields $m_\mathsf{X}$ \cite{BentonThesis}
imposes the conditions 
\begin{eqnarray}
&& 
\frac{2}{\sqrt{3}}
\begin{bmatrix}
\partial_x  m_\mathsf{E}^1  \\
-\frac{1}{2} \partial_y m_\mathsf{E}^1 - \frac{\sqrt{3}}{2} \partial_y m_\mathsf{E}^2  \\
-\frac{1}{2} \partial_y m_\mathsf{E}^1 + \frac{\sqrt{3}}{2} \partial_y m_\mathsf{E}^2
\end{bmatrix}    
- 
\begin{bmatrix}
\partial_y m_\mathsf{T_{1-}}^z + \partial_z m_\mathsf{T_{1-}}^y  \\
\partial_z m_\mathsf{T_{1-}}^x + \partial_x m_\mathsf{T_{1-}}^z  \\
\partial_x m_\mathsf{T_{1-}}^y + \partial_y m_\mathsf{T_{1-}}^x   
\end{bmatrix} 
\nonumber\\
&& -\sqrt{\frac{2}{3}}\bm{\nabla}m_\mathsf{A_{2}}
+ \bm{\nabla} \times \mathbf{m}_\mathsf{T_{2}}
= 0 \; .
\label{eq:HAF.constraint}
\end{eqnarray}
We obtain exactly the same constraint if we 
substitute ${\bf E}^{\sf HAF}$ in Eq.~\eqref{eq:constraint.2}, 
implying that HAF automatically satisfies one of the two constraints
defining the \mbox{R2--U1} spin liquid
\footnote{This system of equations can also be viewed as three 
	independent copies of a $U(1)$ gauge theory \cite{henley05}.}.


To convert the HAF into an \mbox{R2--U1} spin liquid, we need to 
make the theory symmetric and traceless, 
and so satisfy Eq.~\eqref{eq:constraint.1}.
This means eliminating fluctuations of ${\bf E}^{\sf HAF}_ {\sf antisym.}$ 
and ${\bf E}^{\sf HAF}_{\sf trace}$ from the ground state, something 
which can be accomplished by opening gaps to  
the fields $m_\mathsf{T_2}$ and $m_\mathsf{A_2}$.
For the BP 
model, Eq.~(\ref{eq:H}), 
this is achieved by any parameter set for which 
\begin{eqnarray}
J_A\ ,\ J_B>0\ ,\ D_A<0\ ,\ D_B=0 \; . 
\label{eq:master.parameter.set}
\end{eqnarray}
In this case,
the coefficients $a_{\mathsf{X},\text{A}}$  
satisfy the condition
\begin{equation}
a_{\mathsf{E},\text{A}}   =   a_{\mathsf{T_{1-}},\text{A}}  < 
a_{\mathsf{A_2},\text{A}}  ,\  a_{\mathsf{T_2},\text{A} }   , \  a_{\mathsf{T_{1+}},\text{A}} \; ,
\label{eqn.a.in.A}
\end{equation}
which implies that only the fields 
$ \mathbf{m}_{\mathsf{E}} $ and $\mathbf{m}_{\mathsf{T_{1-}}}$ 
enter into the ground state of Eq.~(\ref{eq:H.m.Td}).
Meanwhile, on the B--sublattice, we recover the condition 
Eq.~(\ref{eq:HAF.condition}), previously found for the HAF, which imposes the 
constraint Eq.~(\ref{eq:HAF.constraint}), with the caveat that the fields 
$m_\mathsf{A_{2}}$ and $\mathbf{m}_\mathsf{T_{2}}$ can now be set identically 
equal to zero.
When expressed in terms of the remaining tensor field ${\bf E}^{\sf HAF}_{\sf sym}$, 
this is exactly Eq.~\eqref{eq:constraint.2}.
It follows that, in this classical limit, an \mbox{R2--U1} gauge theory, 
satisfying both Eq.~\eqref{eq:constraint.1} and Eq.~\eqref{eq:constraint.2} 
emerges as the effective description at  the low--energy sector of the BP 
model, Eq.~\eqref{eq:H}.


It is worth noting that, as in the regular pyrochlore lattice \cite{CanalsPRB08,chern2010pyrochlore}, 
DM interaction is only a singular perturbation in the context 
of the classical ground--state manifold.
At finite temperature, classical spin liquids owe their stability to entropy, 
and a finite value of $D_A$ will be needed to stabilise an \mbox{R2--U1} 
spin liquid.
For exactly the same reason, introducing a finite 
value of $D_B$ does not immediately invalidate the mechanism driving
the \mbox{R2--U1} spin liquid, but will reduce the range of 
temperatures over which it is observed.
We will see that both of these expectations are fulfilled by  
classical Monte Carlo simulations of Eq.~(\ref{eq:H}), described below


\begin{figure}[t]
	\centering
	\includegraphics[width=0.8\columnwidth]{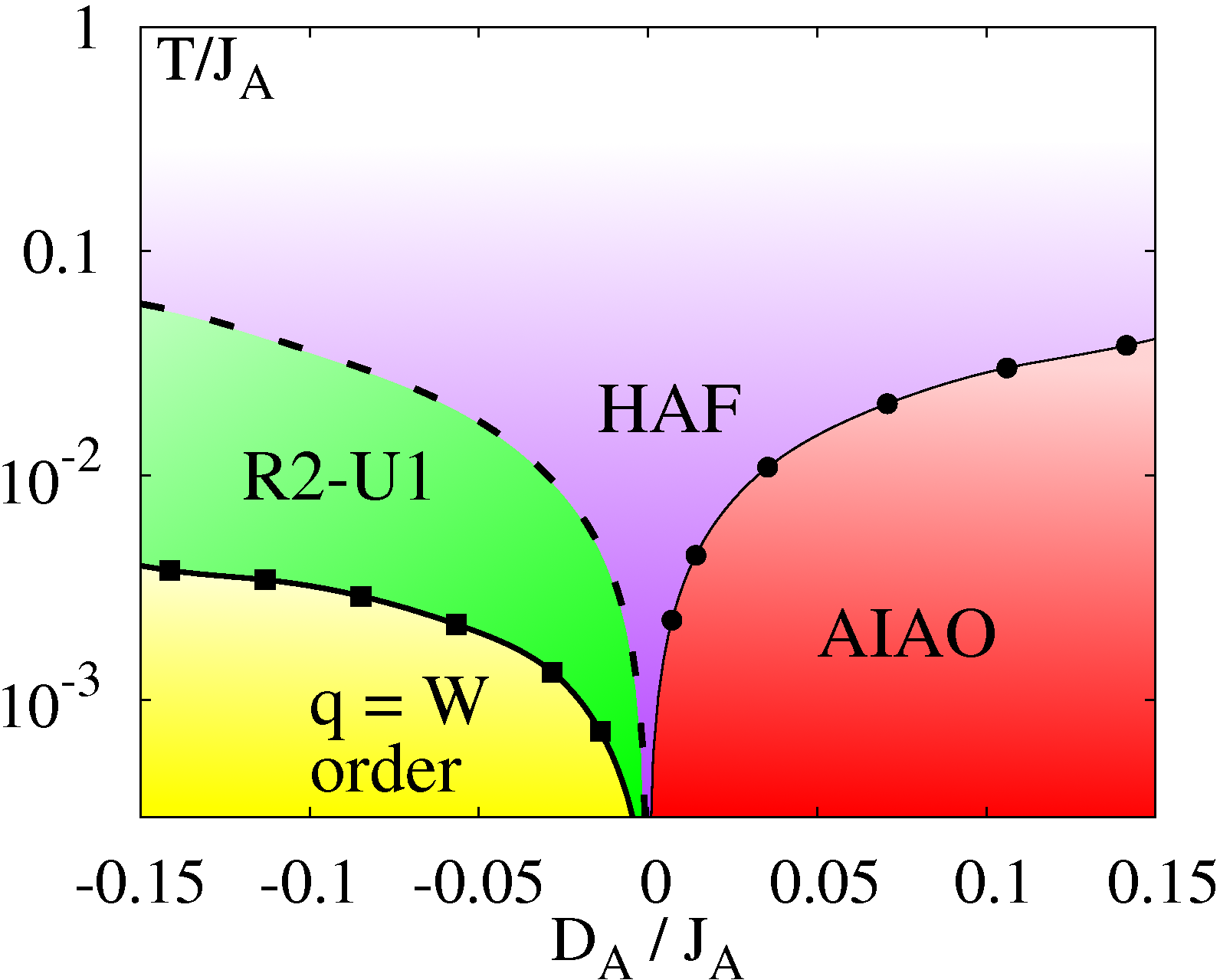}
	\caption{
		%
		%
		%
		Finite--temperature phase diagram of the BP model, 
		Eq.~\eqref{eq:H}, as a function of DM interaction $D_A$.
		The crossover between the \mbox{R2--U1} spin liquid, 
		and the $U(1)\times U(1)\times U(1)$ spin liquid (HAF) 
		is shown with a dashed line.
		The thin solid line indicates a continuous transition into all--in all--out 
		order (AIAO), 
		while thick solid line denotes a first order 
		phase transition into a state with ${\bf q} = W$ order. 
		Results are taken from MC simulation with $J_A = J_B = 1$, $D_B =0$.
	}	\label{fig:phase.diagram}
\end{figure}


\indent
{\it Characteristic signatures of R2--U1 state.}
We now turn to the question of how the R2--U1 spin liquid can be identified, 
in both simulation and in experiment.
The zero--divergence condition in spin ice, Eq.~(\ref{eq:spin.ice.constraint}), 
manifests itself in a pinch--point singularity \cite{henley05}
\begin{eqnarray}
\langle E_i (\bfq) E_j(-\bfq) \rangle 
\propto  
\delta_{ij} -\frac{q_iq_j}{q^2} \; ,
\label{eq:2FPP}
\end{eqnarray}
which is observed in neutron scattering experiments \cite{Fennell2009}.
In the same way, the constraints associated with an \mbox{R2--U1} 
gauge theory, Eq.~(\ref{eq:constraint.1}) and Eq.~(\ref{eq:constraint.2}), lead to a 
characteristic singularity in correlations of the tensor field~$E_{ij}$~\cite{PremPRB18} 
\begin{equation}
\label{eq:4FPP}
\begin{split}
&\langle E_{ij}(\bfq) E_{kl}(-\bfq) \rangle 
\propto  
\frac{1}{2}(\delta_{ik}\delta_{jl} + \delta_{il}\delta_{jk})
+ \frac{q_i q_j q_k q_l}{q^4} 			\\
& \quad - \frac{1}{2} \bigg(\delta_{ik}\frac{q_i q_l}{q^2}
+\delta_{jk}\frac{q_i q_l}{q^2}	
+\delta_{il}\frac{q_j q_k}{q^2}
+\delta_{jl}\frac{q_i q_k}{q^2}	\bigg) 		\\
&\quad  -\frac{1}{2}\left( \delta_{ij} -\frac{q_iq_j}{q^2} \right)\left( \delta_{kl} -\frac{q_kq_l}{q^2} \right) \; .
\end{split}
\end{equation}
The three--dimensional structure of 
the correlation $\langle E_{\sf xy} ({\bf q}) E_{\sf xy} ({\bf -q}) \rangle$
is illustrated in Fig.~\ref{fig:1}.
In the $[0kl]$ plane, correlations exhibit a 
conventional \mbox{2--fold} pinch point, comparable to that found in spin ice [Fig.~\ref{fig:2FPP}].
However in the perpendicular $[hk0]$ plane, we observe 
a 4--fold pinch point (4FPP) [Fig.~\ref{fig:4FPP}], which  
unambiguously distinguishes \mbox{R2--U1} electrodynamics  
from lower--rank theories \cite{PremPRB18}.


{\it Comparison with simulation}. 
We can use the existence of this 4FPP as a test for the
\mbox{R2--U1} spin liquid in simulation.
We have carried out classical Monte Carlo (MC) simulations 
of Eq.~(\ref{eq:H}), for the parameter--set 
\begin{eqnarray}
J_A = J_B = 1\; ,\;  D_A= -0.01\; ,\;  D_B = 0 \; .
\label{eq:idealised.parameter.set}
\end{eqnarray}
where the constraints Eq.~\eqref{eq:constraint.1} and Eq.~\eqref{eq:constraint.2}
are expected to hold.
The resulting correlations of $E_{ij}$, at a temperature $T = 2.5 \times10^{-3}\ J_A$, 
are shown in Fig.~\ref{fig:4FPP-MC}.
For ${\bf q} \to 0$, these are identical to the predictions of Eq.~(\ref{eq:4FPP}), 
confirming that the model realizes an \mbox{R2--U1} spin liquid.


\indent
{\it Phase diagram. }
The results of simulations for a range of values of $D_A$ are collected 
in Fig.~\ref{fig:phase.diagram}.
At finite temperature, a finite value of $D_A < 0$ is required to achieve a crossover 
from the $U(1)\times U(1)\times U(1)$ spin liquid of the pyrochlore HAF, 
with 2--fold pinch points, into an \mbox{R2--U1} spin liquid, with 4FPP.  
An analytic theory of this crossover, which is controlled by the dimensionless 
parameter $\eta \sim |D_A|/k_B T$, is provided in Section VI of 
the Supplemental Material.
Meanwhile, at low temperatures, sufficiently negative values of $D_A$ drive
a first--order phase transition into an ordered state 
which involves the characteristic wavevector $\bfq={\bf W}$ 
(i.e. corners of the Brillouin zone) \footnote{A more complex, 
	multiple--$\bfq$ ground state is hard to rule out categorically, 
	because of the difficulty of thermalising simulations at the 
	lowest temperatures.}.  
In contrast, a finite value of $D_A > 0$ leads to a continuous 
phase transition into a state with $\bfq=0$, all--out (AIAO) order. 



\begin{figure}[t]
	\centering
	\subfloat[$S_{\sf SF}({\bf q})$, $D_B = 0$ \label{fig:Sq.MC.DB.zero}]{\includegraphics[height=0.4\columnwidth]{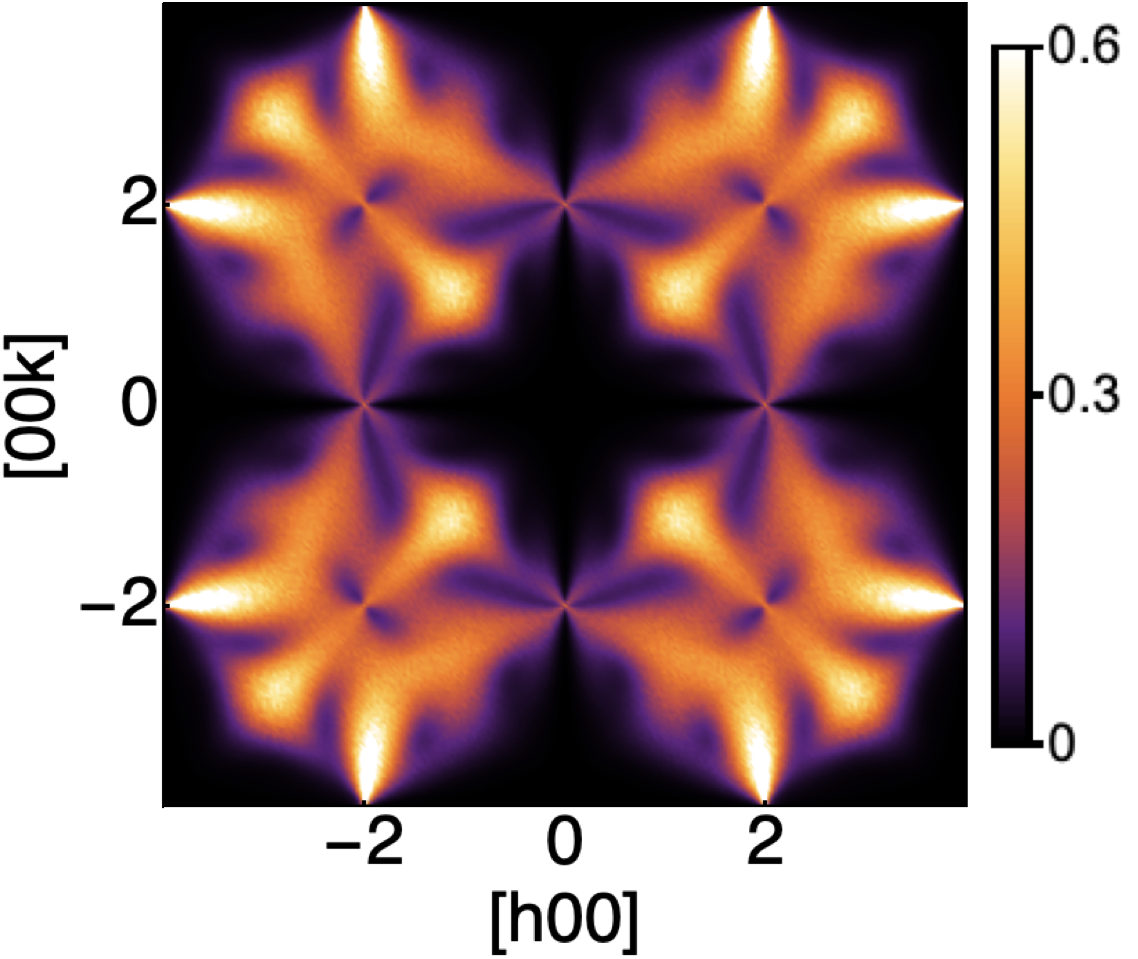}} \;
	\subfloat[$S_{\sf SF}({\bf q})$, $D_B \ll D_A$ \label{fig:Sq.MC.DB.finite}]{\includegraphics[height=0.4\columnwidth]{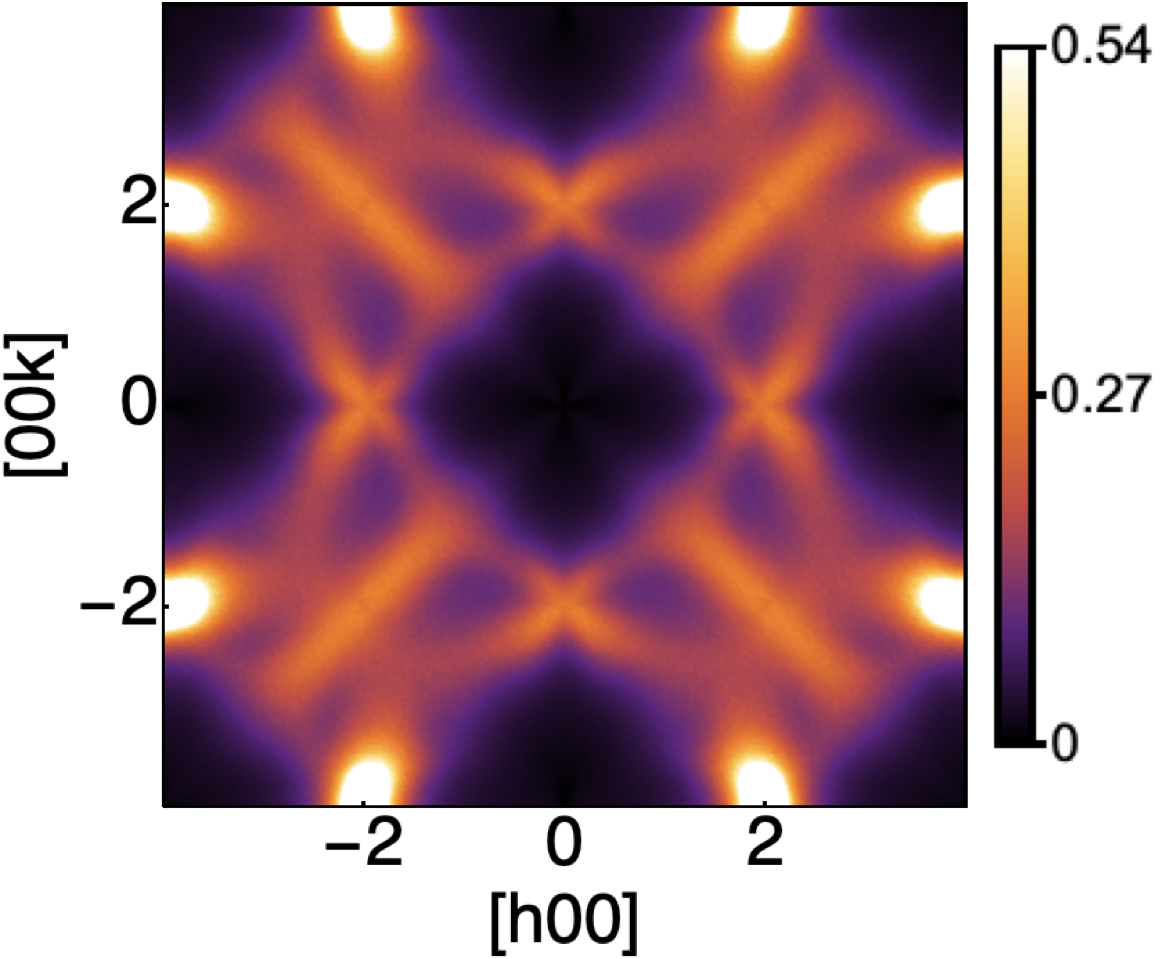}} \\
	\caption{
		Spin structure factor found in MC simulation 
		of the BP model, Eq.~\eqref{eq:H}, 
		showing 4--fold pinch points (4FPPs) characteristic of a 
		\mbox{R2--U1} spin liquid.
		(a) Correlations in the $[h0k]$ plane, in the spin--flip (SF) 
		channel measured using polarised neutrons.
		4FPP are visible at $[0,0,2]$ and points related by symmetry.
		Results are for parameters Eq.~(\ref{eq:idealised.parameter.set}),   
		\mbox{$T =  2.5 \times 10^{-3} J_A$}.	
		(b) Equivalent results for parameters motivated by \BYZO,
		Eq.~(\ref{eq:experimental.parameter.set}), $T =  252\ \text{mK}$.
	}
	\label{fig:Sq.MC}
\end{figure}

\indent
{\it Predictions for neutron scattering.}
%
%
Neutron scattering experiments 
do not measure correlations of $E_{ij}$ directly, but rather the spin structure factor 
\mbox{$S^{\alpha\beta}({\bf q}) = \langle S^\alpha(\bfq) S^\beta(-\bfq) \rangle$}.
On general grounds \cite{PremPRB18},  
$S^{\alpha\beta}({\bf q})$ 
is expected to bear witness to the singularity in Eq.~(\ref{eq:4FPP}).
But exactly how 4FPPs  
would manifest themselves in experiment remains an open question.
In Fig.~\ref{fig:Sq.MC} we present simulation results for $S^{\alpha\beta}({\bf q})$
for parameters equivalent to Fig.~\ref{fig:Sq.MC.DB.zero}.
We find that the 4FPP is {\it not} visible in the structure factor measured
by unpolarised neutron scattering [see Supplemental Material].  
However the 4FPP {\it can} be resolved using polarised neutrons.
In this case, it manifests itself in the spin--flip (SF) channel 
for neutrons polarised perpendicular to the scattering 
plane \cite{Fennell2009}, [Fig.~\ref{fig:4FPP}].


\indent
{\it Application to materials. }
Breathing--pyrochlore magnets were first studied as a tractable limit of the pyrochlore 
HAF \cite{harris91a,canal98-PRL80,canals00-PRB61,tsunetsugu01},  but 
have since been realised in materials based on both transition--metal 
\cite{okamoto13,tanaka14-PRL113,nilsen15-PRB91,wawrzynczak17-PRL119,okamoto18-JPSP87} 
and rare--earth ions \cite{KimuraPRB14,HakuPRB16}. 
Interesting parallels are also found in lacunar spinels \cite{widmann16,jeong17}.   
To date, most theoretical work has concentrated on $SU(2)$--invariant 
models \cite{harris91a,canal98-PRL80,canals00-PRB61,tsunetsugu01,benton15-JPSJ84,li16,essafi17-JPCM29}.
However, in the presence of spin--orbit coupling, the symmetry of the lattice permits anisotropic exchange \cite{RauPRL16,HakuPRB16,Savary16a,Rau2018PRB}.
And, with respect to higher--rank gauge theories, 
a promising line of enquiry are Yb--based materials, 
where the required form of interactions appear to 
predominate.


One concrete example is \BYZO\ \cite{KimuraPRB14,RauPRL16,HakuPRB16,Rau2018PRB}, 
where A--tetrahedra are estimated to have the coupling parameters
\mbox{$J_A \approx 0.57\ \text{meV}$}, 
\mbox{$D_A \approx -0.16\ \text{meV}$},
with other interactions negligible.
This is {\it exactly} the form of interactions needed for an \mbox{R2--U1} spin liquid, 
a feature which is expected to be robust \cite{Rau2018PRB}, since it holds for a 
wide range of Slater--Koster overlap ratios \cite{Slater1954PR}.
Meanwhile, exchange interactions on the larger B--tetrahedra of \BYZO, 
while less well understood, appear to be orders of magnitude smaller \cite{RauPRL16,HakuPRB16}.
Thus, while it seems plausible that \BYZO\ could realise a \mbox{R2--U1} 
spin liquid, this may occur at temperatures too low to measure.   


The encouraging example of \BYZO\ motivates us to consider the possibility 
of a magnet with similar structure, but smaller B--tetrahedra, such that the
interactions on the B--sublattice become non--negligible.  
For concreteness, we consider a parameter set:
\begin{gather}
J_A = 0.57\ \text{meV}, \; 
J_B  = 0.028\ \text{meV} \; , \nonumber\\  
D_A = -0.16\ \text{meV}, \;
D_B =  -0.007\ \text{meV} \; ,
\label{eq:experimental.parameter.set} 
\end{gather}
where we assume that the interactions 
on the B--sublattice are of the same form as on the A--sublattice, 
but substantially weaker, \mbox{$J_A/J_B = D_A/D_B \approx 20$}.
To demonstrate that the \mbox{R2--U1} physics persists in the presence of
finite $D_B$ we have used MC simulation to calculate the spin structure factor.
Once again, the 4FPP associated with the R2--U1 spin liquid remains clearly visible
for a range of temperatures  [Fig.~\ref{fig:Sq.MC.DB.finite}].
The same will hold for a more general choice of interactions, 
as long as the anisotropic part of the exchange on the B--sublattice 
is sufficiently weak; for $D_B\sim D_A$, fluctuations are restricted 
to the local easy plane, and the \mbox{R2--U1} physics will be lost.


\indent
{\it Quantum effects}.  
The theory of an \mbox{R2--U1} spin liquid presented above is 
classical, so it is important to ask what might change once quantum 
effects are taken into account.
A useful point of comparison is quantum spin ice (QSI), 
where quantum fluctuations leads to tunnelling between different spin 
configurations satisfying the ``ice rules'' constraint Eq.~(\ref{eq:spin.ice.constraint}).
This tunnelling, which occurs on loops of spins, introduces a fluctuating 
magnetic field ${\bf B}$, and the result, at $T=0$, 
is a QSL described by a the deconfined phase of a $U(1)$ quantum 
lattice gauge theory \cite{hermele04PRB,banerjee08-PRL100,benton12-PRB86,savary12,
	Shannon2012PRL,hao14,Gingras14RoPP,kato15,chen17,huang18-PRL120}.   
However it is important to note that the temperature scale associated 
with this QSL is three orders of magnitude smaller than the range of 
temperatures over which Eq.~(\ref{eq:spin.ice.constraint}) holds \cite{huang18-PRL120}.
Moreover, since the $U(1)$ QSL is gapless, any finite temperature 
immediately restores classical correlations at long length 
scales \cite{benton12-PRB86}.
As a consequence, the spin structure factor $S({\bf q})$ 
continues to be dominated by pinch--point singularities 
of the form Eq.~(\ref{eq:2FPP}), down to the lowest temperatures 
studied \cite{kato15}.


The quantum limit of R2--U1 gauge theories has already 
been studied as a continuum field theory, and is qualitatively 
very similar to QSI \cite{PretkoPRB16,PretkoPRB17,PremPRB18}.
The lowest lying excitations are gapless emergent photons
which modify, but do not eliminate, the singular features observed 
in scattering \cite{benton12-PRB86,PremPRB18}.
The microscopic study of quantum effects in Eq.~(\ref{eq:H}) 
lies outside the scope of this Letter.
However we anticipate that coherent gauge fluctuations will be confined 
to an even lower temperature scale than in QSI, by the 
fact that the magnetic field $B_{ij}$ is an extended object, involving 
third--order derivatives of $A_{ij}$ \cite{PretkoPRB16,PretkoPRB17}. 
For this reason the classical theory developed here should 
prove sufficient to interpret experiments searching for an \mbox{R2--U1}
in a BP material.


\indent
{\it Summary and perspectives.}  
In this Letter, we have used a combination of analytic field theory
and classical Monte Carlo simulation to show how a \mbox{rank--2} $U(1)$ 
[\mbox{R2--U1}] spin liquid, a state described by a higher--rank 
generalisation of electrodynamics, can arise in a pyrochlore magnet with 
breathing anisotropy and Dzyaloshinskii--Moriya  
interactions, Eq.~(\ref{eq:H}) [cf. Fig.~\ref{fig:4FPP}].
These results provide a concrete starting point for the experimental search 
for higher--rank gauge theories,  
and clarify the type of neutron scattering experiment which would be needed to 
resolve the 4--fold pinch points (4FPP) of a \mbox{R2--U1} spin liquid 
[cf.~Fig.~\ref{fig:Sq.MC}].


This work opens a number of interesting perspectives.
On the experimental side, we identify Yb based breathing--pyrochlore 
materials as potential candidates for a \mbox{R2--U1} spin liquid state.
On the theoretical side, determining the quantum ground state 
of Eq.~(\ref{eq:H}), should ultimately prove tractable, since breathing 
anisotropy provides a natural control parameter for both perturbative 
\cite{canal98-PRL80,canals00-PRB61} and variational 
approaches \cite{benton18-PRL121}.
And, while the model studied here does not correspond to 
a fracton stabilizer code upon Higgsing \cite{BulmashPRB2018,MaPRB2018}, the 
parital--confinement mechanism used to eliminate selected components 
of the tensorial electric field is very versatile, and easily adapted 
to generate other versions of \mbox{R2--U1} theory \cite{Han-Owen-unpublished}.


\indent
{\it Acknowledgements.}
The authors acknowledge helpful conversations with Jeffrey Rau 
and Daniel Khomskii. 
This work was supported by the Theory of Quantum Matter Unit, Okinawa Institute 
of Science and Technology Graduate University (OIST).
H.Y. is also supported by 
Japan Society for the Promotion of Science (JSPS) Research 
Fellowship for Young Scientists. 
L. J. acknowledges financial support from the French ``Agence Nationale de la Recherche'' under Grant No. ANR-18-CE30-0011-01, and hospitality from Gakushuin University under Grants-in-Aid for Scientific Research on innovative areas ``Topological Materials Science'' (No.JP15H05852) from JSPS.
The research was also  supported in part by the National Science Foundation under Grant No. NSF PHY-1748958,
and the KITP program "Topological Quantum Matter: From Concepts to Realizations".

\bibliography{R2U1-letter.bib}



\clearpage
\onecolumngrid

\begin{center}
	\textbf{\large Supplemental Material}
\end{center}



\setcounter{equation}{0}
\setcounter{figure}{0}
\setcounter{table}{0}
\setcounter{page}{1}
\makeatletter
\renewcommand{\theequation}{S\arabic{equation}}
\renewcommand{\thefigure}{S\arabic{figure}}

\section{Rank--2 $U(1)$ gauge theory: electrostatics}

Here, following Ref.~\cite{XuPRB06,PretkoPRB16}, we derive the relationship between electric, magnetic 
and gauge fields within the rank--2 $U(1)$ [\mbox{R2--U1}] electrodynamics considered in the main text.
This section focuses on the classical electrostatics which is realized in our work. 
The next section will focus on the quantum dynamics of the theory, which is beyond the scope of this work but nevertheless 
essential for future developments.

Our starting point is an electric field described by a symmetric, traceless rank--2 tensor,
\begin{equation}
E_{ij} = E_{ji} , \quad  E_{ii} = 0.
\end{equation}
Here we do not distinguish super and subscript since we are dealing with spacial indices.

The low energy sector of the electric field has vanishing vector charge, and is traceless,
\begin{equation}\label{EQN_S_low_E_cond}
\partial_i E_{ij} = 0
\end{equation}
Here we keep all indices as subscripts but still use the Einstein summation rule. 
	The proper rank-2 tensor with the proper Gauss law as a classical spin liquid system 
	is achieved in this work.

\section{Rank--2 $U(1)$ gauge theory: dynamics}

	A quantum spin liquid requires quantum dynamics in addition to the emergent Gauss law.
	Broadly speaking, the dynamics 
	play the role of ${\bf B}^2$ term in electrodynamics.
	They are to tunnel different  classical spin liquid states between each other, 
	leading to a long-range entangled quantum ground state and gapless photon excitations.
	This section explains how to derive such terms and also their implication on mobility of electric charges (fractons).

As in conventional electrodynamics, the conjugate  of ${\bf E}$ is the rank-two gauge field $\bf A$, which
also has to be symmetric to match the degrees of freedom,
\begin{equation}
A_{ij} = A_{ji}.
\end{equation}
%

These two conditions determine the form of gauge transformation. 
Consider a  wave-function
\begin{equation}
\ket{\Psi({\bf A})}.
\end{equation}
We take a low energy configuration of $\bf E$ obeying Eq.~\eqref{EQN_S_low_E_cond} and construct a symmetrized operator that is identical to zero to act upon the wave-function
\begin{equation}
-i(\lambda_j \partial_i E_{ij} + \lambda_j  \partial_i E_{ij})\ket{\Psi({\bf A})} = 0.
\end{equation}
By integration by parts and assuming vanishing boundary terms, we have 
\begin{equation}
i(\partial_i\lambda_j +\partial_i \lambda_j )E_{ij}\ket{\Psi({\bf A})} = 0.
\end{equation}
Since $E_{ij}$ conjugates with $A_{ij}$, it generates a transformation of $\bf A$.
Thus
\begin{equation}
i(\partial_i\lambda_j +\partial_i \lambda_j )E_{ij}\ket{\Psi({\bf A})} = \ket{\Psi({\bf A}+\bm{\nabla} \otimes\bm{\lambda} +(\bm{\nabla} \otimes\bm{\lambda} )^T)}  -   \ket{\Psi({\bf A})}  = 0.
\end{equation}
That is, the low energy sector wave-function is invariant under gauge transformation 
\begin{equation}
\begin{split}
{\bf A}+\bm{\nabla} \otimes\bm{\lambda} +(\bm{\nabla} \otimes\bm{\lambda} )^T, \quad
\text{ i.e., }\quad A_{ij }\rightarrow A_{ij} + \partial_i\lambda_j +\partial_i \lambda_j.
\end{split}
\end{equation}


Similarly, the traceless condition  
\begin{equation}
-i\gamma\delta_{ij} E_{ij}\ket{\Psi({\bf A})} = 0.
\end{equation}
leads to another gauge symmetry
\begin{equation}
A_{ij }\rightarrow A_{ij} +\gamma\delta_{ij}.
\end{equation}


Finally, the magnetic field is obtained by finding the simplest gauge-invariant quantity.
In this case, it has to have three derivatives acting on the gauge field,
\begin{equation}
\begin{split}
B_{ij} = & \frac{1}{2}[ \epsilon_{jab}(\partial_a \partial_k \partial_i A_{bk} - \partial_a \partial^2 A_{bi})		\\
& + \epsilon_{iab} (\partial_a \partial_k \partial_j A_{bk} - \partial_a \partial^2 A_{bj})].
\end{split}
\end{equation}

Finally, the Gauss law, the traceless and symmetric conditions of the electric field can be used to derive:

\begin{eqnarray} 
&&\int dv \vec{\rho}  = 0 \\
&&\int dv \vec{x} \times \vec{\rho} = - \int dv \epsilon_{ijk}E_{jk} = 0 \\
&& \int dv \vec{x} \cdot \vec{\rho}  = - \int dv E_{ii} = 0 
\end{eqnarray}

In this case, a vector charge excitation is fully fractonic, i.e., it cannot move in any direction of the system.

Further details of the phenomenology of \mbox{R2--U1} phases can be found in Refs.~\cite{PretkoPRB16,PretkoPRB17}.

\section{Derivation of Effective Field Theory}

We show how a rank--2 tensor electric field, 
satisfying the constraint required for \mbox{R2--U1} electrodynamics [Eqs.~(1,2)], 
can be derived from a breathing pyrochlore lattice model [Eq.~(6)].
The pattern of this derivation closely follows Refs.~\cite{Benton2016NatComm,BentonThesis,YanPRB17}.


Our starting point is the breathing  pyrochlore lattice
with a spin on each of its sites,
and nearest neighbor interactions between the spins. 
``Breathing" means the lattice is bi-partitioned into 
A- and B-tetrahedra [Fig.~(1)],
and each type of tetrahedron has its own interactions. 


The model 
that hosts a rank-2 spin liquid
has breathing Heisenberg antiferromagnetic interactions
on both the A- and B-tetrahedra,
and negative Dzyaloshinskii-Moriya  (DM) interactions on A-tetrahedra only.
The Hamiltonian for the model is
\begin{equation}\label{EQN_S_hamiltonian}
\mathcal{H} =
\sum_{\langle ij \rangle\in \text{A}} \left[J_A \bfs_i \cdot \bfs_j 
+
D_A\hat{\bf d}_{ij}\cdot( \bfs_i \times \bfs_j )\right ] 
+ 
\sum_{\langle ij \rangle\in \text{B}} \left[J_B \bfs_i \cdot \bfs_j 
+
D_B  \hat{\bf d}_{ij}\cdot( \bfs_i \times \bfs_j )\right ].
\end{equation}
where $\langle ij \rangle \in \text{A(B)}$ denotes
nearest neightbour bonds belonging to the A(B)-tetrahedra.
The sites $0,\ 1,\ 2,\ 3$ are at positions relative to the center of an A-tetrahedron
\begin{equation}
{\bf r}_0 = \frac{a}{8}(1,1,1),\; {\bf r}_1 = \frac{a}{8}(1,-1,-1),\;
{\bf r}_2 = \frac{a}{8}(-1,1,-1),\; {\bf r}_3 = \frac{a}{8}(-1,-1,1),
\end{equation}
where $a$ is the length of the unit cell.
Vectors $\hat{\bf d}_{ij}$ are bond dependent, defined in accordance with Ref~\cite{KotovPRB05,CanalsPRB08,RauPRL16}:
\begin{equation}
\begin{split}
&\hat{\bf d}_{01}= \frac{(0,-1,1)}{\sqrt{2}},\; 	\hat{\bf d}_{02}= \frac{(1,0,-1)}{\sqrt{2}},\;	\hat{\bf d}_{03}= \frac{(-1,1,0)}{\sqrt{2}} , \\
&\hat{\bf d}_{12}= \frac{(-1,-1,0)}{\sqrt{2}},\;	\hat{\bf d}_{13}= \frac{(1,0,1)}{\sqrt{2}},\;	\hat{\bf d}_{23}= \frac{(0,-1,-1)}{\sqrt{2}}.
\end{split}
\label{EQN_SM_D_Vec}
\end{equation}



Equivalently, 
this model can be written in a standard matrix-exchange form 
for a breathing-pyrochlore lattice model as 
\begin{equation}
\mathcal{H} = 
\sum_{\langle ij \rangle \in \text{A}} S_i^\alpha \mathcal{J}_\text{A,ij}^{\alpha\beta} S_j^\beta
+
\sum_{\langle ij \rangle \in \text{B}} S_i^\alpha \mathcal{J}_\text{B}^{\alpha\beta} S_j^\beta
\end{equation}
where $\mathcal{J}_\text{A,ij}$ is a three-by-three matrix that couples 
spins on sub-lattice sites $i, j$ whose bond belongs to A-tetrahedra,
and  $\mathcal{J}_\text{B}$ is the coupling matrix for B-tetrahedra.
In the case of $D_B=0$ that we are interested in,
$\mathcal{J}_\text{B} $ is identical for any pair of $i, j$,
\begin{equation}
\mathcal{J}_\text{B} = 
\begin{bmatrix}
J_B   &  0 & 0 \\
0 &   J_B    &  0  \\
0   &   0   &  J_B
\end{bmatrix}		.
\end{equation}
Matrices $\mathcal{J}_\text{A,ij}$ are bond dependent and related to each other by the lattice symmetry.
Their values are
\begin{equation}
\begin{split}
&
\mathcal{J}_\text{A,01} = 
\begin{bmatrix}
J_A   &  D_A/\sqrt{2} & D_A/\sqrt{2} \\
-D_A/\sqrt{2} &   J_A    &  0  \\
-D_A/\sqrt{2}   &   0   &  J_A
\end{bmatrix}		,\;
\mathcal{J}_\text{A,02} = 
\begin{bmatrix}
J_A   &  -D_A/\sqrt{2} & 0 \\
D_A/\sqrt{2} &   J_A    &  D_A/\sqrt{2}  \\
0   &  -D_A/\sqrt{2}   &  J_A
\end{bmatrix}		,\; \\
&
\mathcal{J}_\text{A,03} = 
\begin{bmatrix}
J_A   & 0 & -D_A/\sqrt{2}\\
0 &   J_A    &  -D_A/\sqrt{2}  \\
D_A/\sqrt{2}   &  D_A/\sqrt{2}  &  J_A
\end{bmatrix}		,\;	
\mathcal{J}_\text{A,12} = 
\begin{bmatrix}
J_A   &  0 & D_A/\sqrt{2} \\
0 &   J_A    &  -D_A/\sqrt{2}  \\
-D_A/\sqrt{2}   &  D_A/\sqrt{2}  &  J_A
\end{bmatrix}		,\;\\
&
\mathcal{J}_\text{A,13} = 
\begin{bmatrix}
J_A   & D_A/\sqrt{2} & 0\\
-D_A/\sqrt{2} &   J_A    & D_A/\sqrt{2}  \\
0  &  -D_A/\sqrt{2}   &  J_A
\end{bmatrix}		,\;
\mathcal{J}_\text{A,23} = 
\begin{bmatrix}
J_A   & -D_A/\sqrt{2} & D_A/\sqrt{2}\\
D_A/\sqrt{2} &   J_A    & 0  \\
-D_A/\sqrt{2}  &  0   &  J_A
\end{bmatrix}		.
\end{split}
\end{equation}


\begin{table*}
	\begin{tabular}{  c  c   c  }
		\hline
		\hline
		\multirow{2}{*}{}
		order  & 
		definition in terms   & 
		associated
		\\
		parameter & 
		of spin components & 
		ordered phases 
		\\
		\hline
		\multirow{1}{*}{}
		$m_{\sf A_2}$ & 
		$\frac{1}{2 \sqrt{3} } 
		\left(S_0^x+S_0^y+S_0^z+S_1^x-S_1^y-S_1^z-S_2^x+S_2^y-S_2^z-S_3^x-S_3^y+S_3^z
		\right)$ & 
		``all in-all out'' 
		\\   
		\multirow{1}{*}{} 
		${\bf m}_{\sf E}$ & 
		$\begin{pmatrix}
		\frac{1}{2 \sqrt{6} } \left( -2 S_0^x + S_0^y + S_0^z - 2 S_1^x - S_1^y-S_1^z+2 S_2^x + S_2^y-
		S_2^z +2 S_3^x-S_3^y +S_3^z \right) \\
		\frac{1}{2 \sqrt{2}} \left( -S_0^y+S_0^z+S_1^y-S_1^z-S_2^y-S_2^z+S_3^y+S_3^z \right)
		\end{pmatrix}$ &
		$\begin{matrix} \Gamma_{5}, \textrm{including}\\ \Psi_2 \textrm{ and } \Psi_3 \end{matrix}$ \\ 
		\multirow{1}{*}{}
		${\bf m}_{\sf T_{1+}}$  & 
		$\begin{pmatrix}
		\frac{1}{2} (S_0^x+S_1^x+S_2^x+S_3^x) \\
		\frac{1}{2} (S_0^y+S_1^y+S_2^y+S_3^y) \\
		\frac{1}{2} (S_0^z+S_1^z+S_2^z+S_3^z)
		\end{pmatrix} $ &
		collinear FM
		\\
		\multirow{1}{*}{}
		${\bf m}_{\sf T_{1, -}}$  & 
		$\begin{pmatrix}
		\frac{-1}{2\sqrt{2}} (S_0^y+S_0^z-S_1^y-S_1^z-S_2^y+S_2^z+S_3^y-S_3^z)  \\
		\frac{-1}{2\sqrt{2}} (S_0^x+S_0^z-S_1^x+S_1^z-S_2^x-S_2^z+S_3^x-S_3^z)  \\
		\frac{-1}{2\sqrt{2}} ( S_0^x+S_0^y-S_1^x+S_1^y+S_2^x-S_2^y-S_3^x-S_3^y) 
		\end{pmatrix}$ &
		non-collinear FM 
		\\
		\multirow{1}{*}{}
		${\bf m}_{\sf T_2} $ & 
		$\begin{pmatrix}
		\frac{1}{2 \sqrt{2}} 
		\left(
		-S_0^y+S_0^z+S_1^y-S_1^z+S_2^y+S_2^z-S_3^y-S_3^z
		\right) 
		\\
		\frac{1}{2 \sqrt{2}} 
		\left(
		S_0^x-S_0^z-S_1^x-S^z_1-S_2^x+S_2^z+S_3^x+S_3^z
		\right) \\
		\frac{1}{2 \sqrt{2} }
		\left(
		-S_0^x+S_0^y+S_1^x+S_1^y-S_2^x-S_2^y+S_3^x-S_3^y
		\right)
		\end{pmatrix} $  &
		Palmer--Chalker ($\Psi_4$)     
		\\ 
		\hline
	\end{tabular}
	\caption{
		Order parameters ${\bf m}_\mathsf{X}$, describing how the point-group symmetry of a 
		single tetrahedron within the pyrochlore lattice is broken by magnetic order.    
		Order parameters transform according to irreducible representations of the point-group 
		${\sf T}_d$, and are expressed in terms of linear combinations of spin-components 
		${\bf S}_i = (S^x_i, S^y_i, S^z_i)$, 
		in the global frame of the crystal axes --- cf. $\mathcal{H}$~[Eq.~\eqref{EQN_S_hamiltonian})].   
		Labelling of spins within the tetrahedron follows the convention of 
		Ross~{\it et al.}~\cite{RossPRX11}.
		The notation $\Psi_i$ for ordered phases is taken from~Ref.~\cite{PooleJPCM07}.
	}
	\label{table:m_lambda_global}
\end{table*}


The spin degrees of freedom on each tetrahedron
can be rewritten in terms of fields forming the irreducible representations of the lattice symmetry, 
\begin{equation}\label{EQN_S_Irreps}
{m}_{\mathsf{A_2}} ,\quad
\mathbf{m}_{\mathsf{E}} ,\quad
\mathbf{m}_{\mathsf{T_2}} ,\quad
\mathbf{m}_{\mathsf{T_{1+}}} ,\quad 
\mathbf{m}_{\mathsf{T_{1-}}} ,
\end{equation}
whose definition can be found in Table \ref{table:m_lambda_global}.
They are  linear combinations of the spin degrees of freedom, allowing for a fully quadratic Hamiltonian:
\begin{equation}
\mathcal{H}=\frac{1}{2}\sum_{\mathsf{X}}a_{\mathsf{X},\text{A}} m^2_{\mathsf{X},\text{A}} + 
\frac{1}{2}\sum_{\mathsf{X}}a_{\mathsf{X},\text{B}} m^2_{\mathsf{X},\text{B}} ,
\label{eqn:general_ham}
\end{equation}
where $\mathsf{X}$ runs over irreps of the group $T_d$, i.e. $\{ \mathsf{A_2}, \mathsf{E}, \mathsf{T_2}, \mathsf{T_{1+}}, \mathsf{T_{1-}} \}$ as listed in Eq.~\eqref{EQN_S_Irreps},
and the subscript A,B denotes on which type of tetrahedra  they are defined.
The coefficients $a_{\mathsf{X}}$ are listed in Table.~\ref{table:coefficients}.

\begin{table*}[!t]
	\begin{tabular}{  c  c  }
		\hline
		\hline
		coefficient   of $|{\bf m}_{\mathsf{X}}|^2$ in Eq.\eqref{eqn:general_ham} $\qquad$ & 
		definition in terms of  $J$ and $D$ \\
		\hline
		$a_{\mathsf{ A_2}}$ & $ -J- 4D/\sqrt{2} $  \\
		$a_{\mathsf {E}}$ & $ -J + 2 D/\sqrt{2}  $  \\
		$a_{\mathsf{ T_2}}$ & $ -J -2D/\sqrt{2}$  \\
		$a_{\mathsf {T_+}}$ & $3J$  \\
		$a_{\mathsf {T_-}}$ & $-J + 2 D/\sqrt{2}$  \\
		\hline
	\end{tabular}
	\caption{Coefficients $a_{\mathsf{X}}$ of the irrep invariants $|{\bf m}_{\mathsf{X}}|^2$ 
		appearing in $\mathcal{H}$~[Eq.\eqref{eqn:general_ham}].
		Coefficients are expressed as a function of $J$ and $D$.
		Here the subscripts for the A- and B-tetrahedra are suppressed.
	}
	\label{table:coefficients}
\end{table*}

For the couplings in Eq.~\eqref{EQN_S_hamiltonian}, we have on A-tetrahedra
\begin{eqnarray}
a_{\mathsf{A_2},\text{A}}   &  =   &     -J_A- 4D_A/\sqrt{2} \;,\\
a_{\mathsf{T_2},\text{A}}   &  =   &     -J_A -2D_A/\sqrt{2} \;,\\
a_{\mathsf{T_{1+}},\text{A}}   &  =   &   3J_A \;,\\
a_{\mathsf{T_{1-}},\text{A}}  = a_{\mathsf{E},\text{A}} &  =   &   -J_A + 2 D_A/\sqrt{2}  ,
\end{eqnarray}
and on B-tetrahedra
\begin{eqnarray}
a_{\mathsf{A_2},\text{B} }=  a_{\mathsf{E},\text{B}}  = a_{\mathsf{T_2},\text{B}}  = a_{\mathsf{T_{1-}},\text{B} } &  =   &     -J_B    ,\\
a_{\mathsf{T_{1+}},\text{B} }   &  =   &  3J_B  .
\end{eqnarray}
For $J_{A},J_{B}>0$ and $D_{A}<0$, these parameters are in order
\begin{eqnarray}
&& \text{on A-tetrahedra:}  \qquad 
a_{\mathsf{E},\text{A}}   =   a_{\mathsf{T_{1-}},\text{A}}  < 
a_{\mathsf{A_2},\text{A}}  ,\  a_{\mathsf{T_2},\text{A} }   , \  a_{\mathsf{T_{1+}},\text{A}} ,\\
&& \text{on B-tetrahedra:}  \qquad 
a_{\mathsf{A_2},\text{B}} = a_{\mathsf{E},\text{B}} = a_{\mathsf{T_2},\text{B}} =a_{\mathsf{T_{1-}},\text{B}}  < a_{\mathsf{T_{1+}},\text{B}} ,
\end{eqnarray}
which plays the central role of dictating the low energy physics.


The irreducible representation fields are subject to constraints arising from fixed spin length
\begin{equation}
\sum_\mathsf{X} m^2_\mathsf{X} = 1
\end{equation}
for both A- and B-tetrahedra.
As a consequence, the low energy sector allows the $m^2_\mathsf{X}$ corresponding 
to the smallest $a_{\mathsf{X}}$ to fluctuate,
while all other fields have to vanish. 
This principle applied to our model leads to
\begin{itemize}
	\item On A-tetrahedra, the fields $ \mathbf{m}_{\mathsf{E}} $ and $\mathbf{m}_{\mathsf{T_{1-}}}$ can fluctuate;
	\item On A-tetrahedra, the fields $\bfm_{\mathsf{T_{1+}}} = \bfm_{\mathsf{T_2}} = \mathbf{0},\  {m}_{\mathsf{A_2}}=0$;
	\item On B-tetrahedra, the fields ${m}_{\mathsf{A_2}} ,\;  \mathbf{m}_{\mathsf{E}} ,\;
	\mathbf{m}_{\mathsf{T_2}} ,\; \mathbf{m}_{\mathsf{T_{1-}}}$ can fluctuate;
	\item On B-tetrahedra,
	\begin{equation}\label{EQN_S_B_condition}
	\bfm_{\mathsf{T_{1+}}} = 0
	\end{equation}
\end{itemize}


Since every spin is shared by  an A- and a B-tetrahedron, 
the fluctuating fields  $ \mathbf{m}_{\mathsf{E}} $ and $\mathbf{m}_{\mathsf{T_{1-}}}$
on A-tetrahedra must obey additional constraints to respect the
the low-energy sector condition
on B-tetrahedron imposed by Eq.~\eqref{EQN_S_B_condition}.
Assuming that the fields are varying slowly in space 
such that the continuous limit can be taken,
the constraint Eq.~\eqref{EQN_S_B_condition}
can be expressed in terms of fields living on A-tetrahedron as
\begin{equation}\label{EQN_S_E_constraint_1}
\frac{2}{\sqrt{3}}
\begin{bmatrix}
\partial_x  m_\mathsf{E}^1  \\
-\frac{1}{2} \partial_y m_\mathsf{E}^1 - \frac{\sqrt{3}}{2} \partial_y m_\mathsf{E}^2  \\
-\frac{1}{2} \partial_y m_\mathsf{E}^1 + \frac{\sqrt{3}}{2} \partial_y m_\mathsf{E}^2
\end{bmatrix}    
- 
\begin{bmatrix}
\partial_y m_\mathsf{T_{1-}}^z + \partial_z m_\mathsf{T_{1-}}^y  \\
\partial_z m_\mathsf{T_{1-}}^x + \partial_x m_\mathsf{T_{1-}}^z  \\
\partial_x m_\mathsf{T_{1-}}^y + \partial_y m_\mathsf{T_{1-}}^x   
\end{bmatrix}  = 0   .
\end{equation}

From this constraint we can build the symmetric, traceless, rank-two magnetic field
$E_{ij}$ as
\begin{equation}
E_{ij} = 
\begin{bmatrix}
\frac{2}{\sqrt{3}}m_\mathsf{E}^1  &  m_\mathsf{T_{1-}}^z &  m_\mathsf{T_{1-}}^y  \\
m_\mathsf{T_{1-}}^z  & -\frac{1}{\sqrt{3}}m_\mathsf{E}^1 - m_\mathsf{E}^2  &   m_\mathsf{T_{1-}}^x  \\
m_\mathsf{T_{1-}}^y &  m_\mathsf{T_{1-}}^x   &  -\frac{1}{\sqrt{3}}m_\mathsf{E}^1 + m_\mathsf{E}^2
\end{bmatrix} \;  ,
\label{eq:S.31}
\end{equation}
such that Eq.~\eqref{EQN_S_E_constraint_1} becomes
\begin{equation}
\partial_i E_{ij}= 0 \; , 
\label{EQN_S_R2_Constraint}
\end{equation}
with symmetric and traceless conditions
\begin{equation} \label{EQN_S_S_E_constaint}
E_{ji}  = E_{ji} ,\qquad  \Tr {\bf E} = 0 
\end{equation}
by the definition of $E_{ij}$.
Hence a rank-2, traceless, vector charged magnetic field emerges 
at the low-energy sector from the microscopic model of Eq.~\eqref{EQN_S_hamiltonian}.


Equation~\eqref{EQN_S_S_E_constaint} constrains the form of correlations functions of
$\langle E_{ij}(\bfq) E_{kl}(-\bfq) \rangle$, in the same spirit as how the two-in-two-out condition constrains the spin-spin correlation of spin ice.
It is, however,  in a more  complicated form.
The explicit results for the \textit{traceful} scalar-charged and vector-charged 
versions of \mbox{R2--U1} are discussed in detail in Ref.~\cite{PremPRB18}.
The vector-charge field correlation is 
\begin{equation}\label{EQN_S__Vc_Corr}
\begin{split}
\langle E_{ij}(\bfq) E_{kl}(-\bfq) \rangle 
\propto  
& \frac{1}{2}(\delta_{ik}\delta_{jl} + \delta_{il}\delta_{jk})
+ \frac{q_i q_j q_k q_l}{q^4} 			\\
& - \frac{1}{2} \left(\delta_{ik}\frac{q_i q_l}{q^2}
+\delta_{jk}\frac{q_i q_l}{q^2}
+\delta_{il}\frac{q_j q_k}{q^2}
+\delta_{jl}\frac{q_i q_k}{q^2}	\right) 
\end{split}
\end{equation}
In close analogy, we derive the correlation function of our \textit{traceless}  vector-charged 
model by deducting the trace,
\begin{equation}\label{EQN_S_Corr}
\begin{split}
\langle E_{ij}(\bfq) E_{kl}(-\bfq) \rangle 
\propto  
& \frac{1}{2}(\delta_{ik}\delta_{jl} + \delta_{il}\delta_{jk})
+ \frac{q_i q_j q_k q_l}{q^4} 			\\
& - \frac{1}{2} \left(\delta_{ik}\frac{q_i q_l}{q^2}
+\delta_{jk}\frac{q_i q_l}{q^2}
+\delta_{il}\frac{q_j q_k}{q^2}
+\delta_{jl}\frac{q_i q_k}{q^2}	\right) 		\\
& -\frac{1}{2}\left( \delta_{ij} -\frac{q_iq_j}{q^2} \right)\left( \delta_{kl} -\frac{q_kq_l}{q^2} \right)	,
\end{split}
\end{equation}
which encodes a singularity at $\bfq \rightarrow 0$.
Different choices of the components $E_{ij}$ and $E_{kl}$ show different patterns.
A few representative ones can be found in Figs.~\ref{fig:4FPP},\ref{Fig_S_diff_corr}.


\begin{figure*}[t]
	\centering
	\subfloat[\label{fig:4FPP1}]{\includegraphics[height=0.2\textwidth]{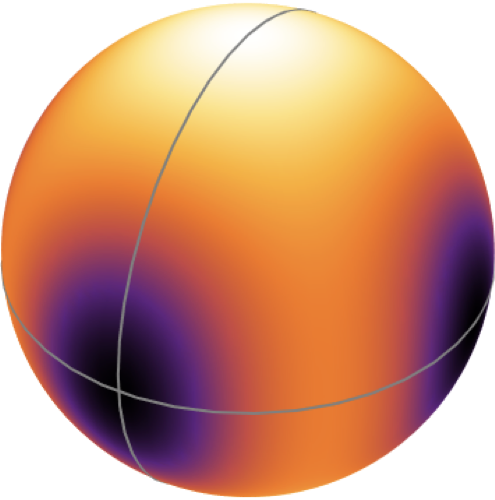}} 
	\qquad
	\subfloat[\label{fig:4FPP2}]{\includegraphics[height=0.2\textwidth]{4FPP_2.png}} 
	\qquad
	\subfloat[\label{fig:4FPP3}]{\includegraphics[height=0.2\textwidth]{4FPP_3.png}} 
	\caption{
		Structure of the 4--fold pinch point (4FPP) associated with rank--2 $U(1)$ [\mbox{R2--U1}] 
		gauge theory.
		(a) Prediction of \mbox{R2--U1} theory for the correlation function 
		$\langle E_{\sf xy} ({\bf q}) E_{\sf xy} ({\bf -q}) \rangle$, on a surface of fixed 
		$|{\bf q}|$ near to a Brillouin zone center.
		(b) Exploded view, showing a 2--fold pinch point in the [0kl] plane.
		(c) Perpendicular cut, showing a 4FPP in the [hk0] plane.
	}
	\label{fig:4FPP}
\end{figure*}


\begin{figure}[H]
	\centering
	\subfloat[\label{EQN_S_Corr_1}]{\includegraphics[width=0.18\textwidth]{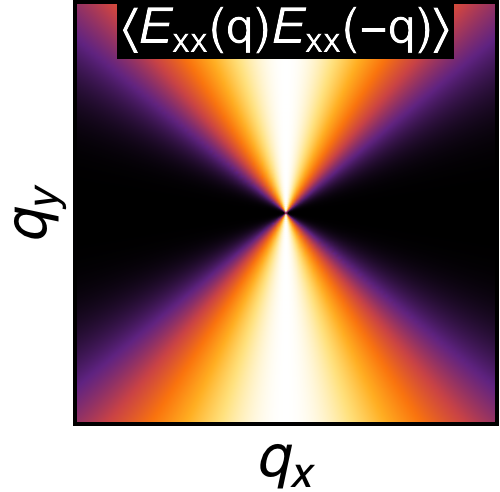}} \;
	{\raisebox{3ex}{\includegraphics[height=0.15\textwidth]{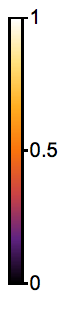}}}\;
	\subfloat[\label{EQN_S_Corr_2}]{\includegraphics[width=0.18\textwidth]{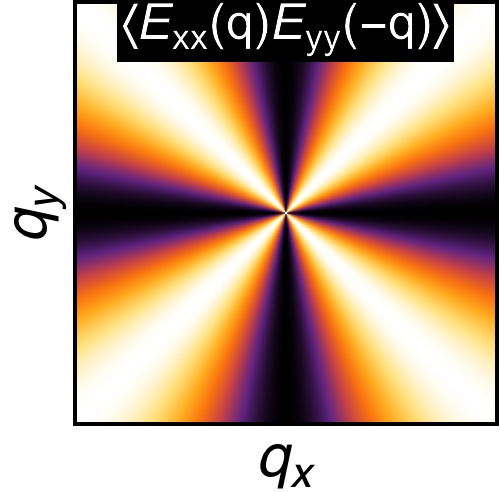}}\;
	\subfloat[\label{EQN_S_Corr_3}]{\includegraphics[width=0.18\textwidth]{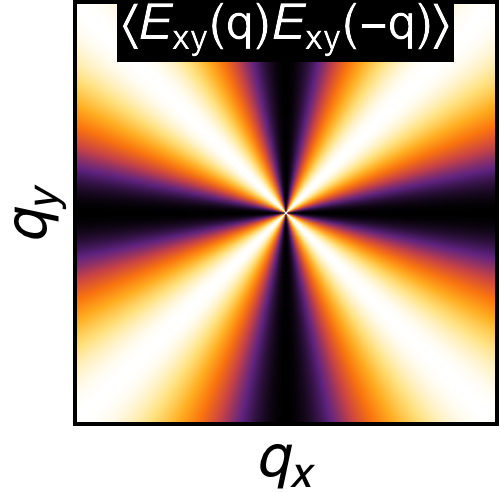}} \;
	{\raisebox{3ex}{\includegraphics[height=0.15\textwidth]{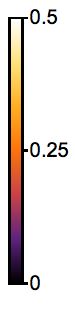}}}\;
	\subfloat[\label{EQN_S_Corr_4}]{\includegraphics[width=0.18\textwidth]{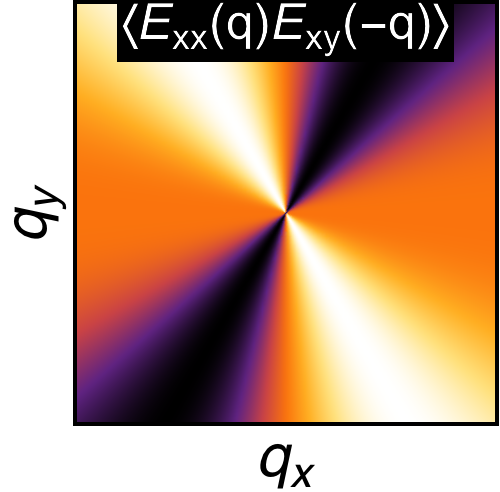}}\;
	\subfloat{\raisebox{3ex}{\includegraphics[height=0.15\textwidth]{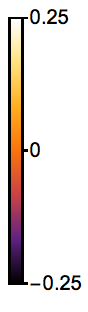}}}
	\caption{Different components of correlation function $\langle E_{ij}(\bfq)E_{kl}(-\bfq)\rangle$
		in $q_x-q_y$ plane,  calculated from Eq.~\eqref{EQN_S_Corr}.
	} 
	\label{Fig_S_diff_corr}
\end{figure}


Fig.~\ref{EQN_S_Corr_2},\ref{EQN_S_Corr_3} have the four-fold pinch-point (4FPP) singularity,
which   differentiates the rank-2 gauge theories uniquely from the conventional $U(1)$ gauge theory.
It is the key signature to be looked for in experiments.
%

\section{Alternative forms of R2--U1 spin liquid}

	In the main text we have shown how a classical spin liquid described 
	by a symmetric, traceless \mbox{R2--U1} gauge theory, descends from 
	the $U(1)\times U(1)\times U(1)$ spin liquid found in the classical 
	Heisenberg Antiferromagnet (HAF) on a pyrochlore lattice.    
	The approach taken is very versatile, and by tuning the Hamiltonian, 
	one can also obtain other forms of \mbox{R2--U1} spin liquid.

	Notice that the diagonal and off-diagonal components of ${\bf E}^{\sf HAF}_{\sf sym.}$
	come from different irreps ${\bf m}_\mathsf{E}$ and ${\bf m}_\mathsf{T_{1-}}$.
	In the most general Hamiltonian (Eq.\ref{eqn:general_ham}), 
	these two irreps have their individually tunable coefficients $a_\mathsf{E}$ and $a_\mathsf{T_{1-}}$.
	So the symmetric part of ${\bf E}^{\sf HAF}$ can be decomposed into three components 
	\begin{equation}
	{\bf E}^{\sf HAF}_{\sf trace}  + {\bf E}^{\sf HAF}_{\sf sym - diagonal} + {\bf E}^{\sf HAF}_{\sf symm-off- diagonal},
	\end{equation}
	and each component can be individually tuned to be active or suppressed by choosing the proper Hamiltonian. 
	The vector-charged Gauss law is unaffected.
	This allows us to build a variety of rank-2 U(1) gauge theories, including a ``hollow'' 
	version with vanishing diagonal terms \cite{MaPRB2018} .

\section{Predictions for Neutron Scattering}

The 4FPP is a unique pattern that differentiates the \mbox{R2--U1} from 
vector $U(1)$ gauge theory, which only has the conventional two-fold pinch points.
The 4FPPs are most unambiguously presented in the correlation function of
the irrep fields as discussed in the main text.
These correlation functions are, however,
not directly accessible in experiment.


In magnetism, neutron scattering is widely used
to measure the spin-spin correlation of the form
\begin{equation}
S(\bfq) = \sum_{\alpha, \beta, i, j} \left( \delta_{\alpha\beta}-\frac{q^\alpha q^\beta}{q^2} \right)
\langle S_i^\alpha(\bfq) S_j^\beta(-\bfq)
\rangle
\end{equation}
where $\alpha, \beta = x, y, z$ are spin-component indices and 
$i, j = 0, 1, 2, 3$ are sub-lattice site indices.


Furthermore, with neutrons polarized in direction of unit vector $\hat{\bf v}$ 
perpendicular to the scattering plane,
one can measure
the spin-flip (SF) channel neutron scattering defined by
\begin{equation}
S(\bfq)_\text{SF} = \sum_{\alpha, \beta, i, j} (v_\perp^\alpha v_\perp^\beta)
\langle S_i^\alpha(\bfq) S_j^\beta(-\bfq)
\rangle		,
\end{equation}
where
\begin{equation}
\hat{\bf v}_\perp = \frac{\hat{\bf v}\times \bfq}{|\hat{\bf v}\times \bfq|} .
\end{equation}
One can also measure the non-spin-flip (NSF) channel defined by
\begin{equation}
S(\bfq)_\text{NSF} =  \sum_{\alpha, \beta, i, j} (v^\alpha v^\beta)
\langle S_i^\alpha(\bfq) S_j^\beta(-\bfq)
\rangle
\end{equation}




Here we show the spin structure factor of the $[h0k]$ and $[hhk]$ plane, with zoomed-in view of the 4FPPs.


\begin{figure}[h]
	\includegraphics[width=0.95\textwidth]{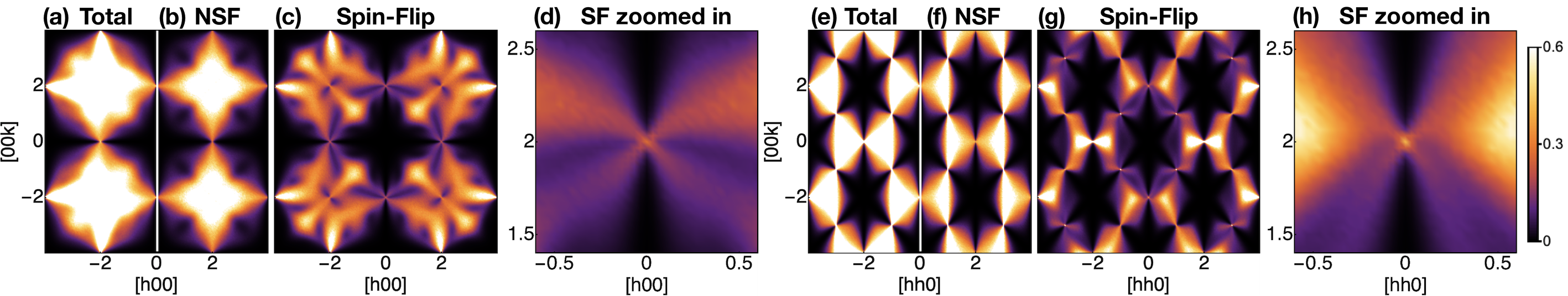}
	\caption{4-Fold Pinch Points (4FPPs)
		in spin structure factor in the $[h0k]$ and $[hhk]$ plane of momentum space of the model  [Eq.~(6)] from MC simulations. The exchange parameters are from the idealized theoretical case $J_A = J_B = 1.0,\ D_A = -0.01,\ D_B = 0.0$, at $T = 2.5 \times10^{-3}\ J_A$.
		(a) Total structure factor in  $[h0k]$ plane. 
		(b) Non-spin-flip (NSF) channel in  $[h0k]$ plane. 
		(c) Spin-flip (SF) channel in  $[h0k]$ plane. 
		(d) Enlarged 4FPP in  $[h0k]$ plane.
		(a) Total structure factor in  $[hhk]$ plane. 
		(b) Non-spin-flip (NSF) channel in  $[hhk]$ plane. 
		(c) Spin-flip (SF) channel in  $[hhk]$ plane. 
		(d) Enlarged 4FPP in  $[hhk]$ plane.
		The 4FPPs can be clearly observed in the SF channel, centered on [0, 0, 2] (and points related by symmetry), but weaker than in the $[h0k]$ plane.	}
	\label{Fig_thy_SF_hhk}
\end{figure}


\begin{figure}[h]
	\includegraphics[width=0.95\textwidth]{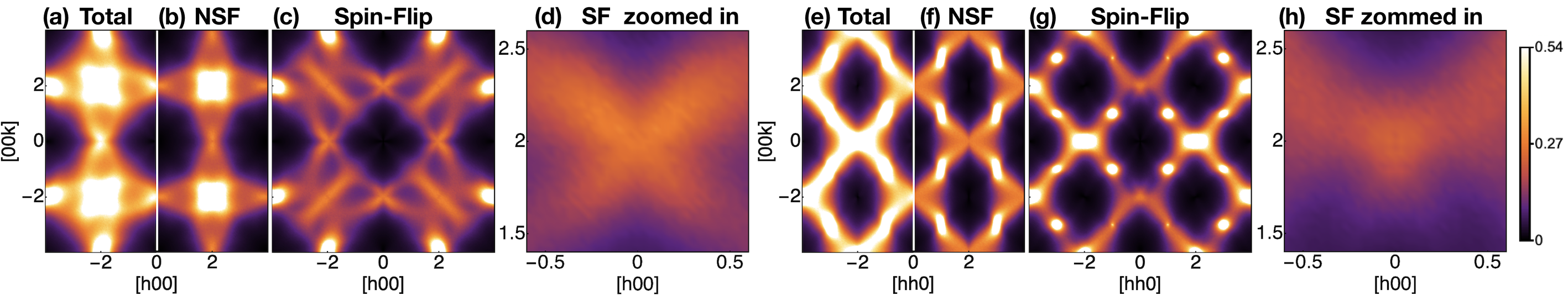}
	\caption{4-Fold Pinch Points (4FPPs)
		in spin structure factors in the $[h0k]$ and $[hhk]$ planes of momentum space of the model  [Eq.~(6)] from MC simulations. The exchange parameters are from the experimental case Eq.~(19) at  $T =  252\ \text{mK}$. 
		(a) Total structure factor in  $[h0k]$ plane. 
		(b) Non-spin-flip (NSF) channel in  $[h0k]$ plane. 
		(c) Spin-flip (SF) channel in  $[h0k]$ plane. 
		(d) Enlarged 4FPP in  $[h0k]$ plane.
		(a) Total structure factor in  $[hhk]$ plane. 
		(b) Non-spin-flip (NSF) channel in  $[hhk]$ plane. 
		(c) Spin-flip (SF) channel in  $[hhk]$ plane. 
		(d) Enlarged 4FPP in  $[hhk]$ plane.
		The 4FPPs can be observed in the SF channel, centered on [0, 0, 2] (and points related by symmetry), but weaker than in the $[h0k]$ plane.	}
	\label{fig:Sq.experimental.parameters_hhk}
\end{figure}

\newpage

\section{Temperature evolution of 4--fold pinch point into 
		a conventional 2--fold pinch point}

The cross--over from the HAF phase to 
	the R2--U1 phase, with decreasing temperature, is manifested in the structure factor as a 
	cross--over between a conventional, 2--fold, pinch point and the 4--fold pinch point (4FPP)
	characteristic of an R2--U1 gauge theory \cite{PremPRB18}.
To provide further quantitative details of this cross--over,
	here we present an analysis of correlations based on a coarse--grained field theory.


As shown above, the $U(1)\times U(1) \times U(1)$ spin liquid of the Heisenberg 
model can be described in terms of a non-symmetric matrix ${\bf E}$, with a Gauss'
law applied to each column.
The traceless R2-U(1) spin liquid is described in the same way, but with the
constraint that ${\bf E}$ be symmetric and traceless.

The following effective theory captures both cases:
\begin{eqnarray}
&&\mathcal{Z}=\int \prod_{\mu,\nu} dE_{\mu\nu}
\exp(-\beta H_{eff}[E_{\mu \nu}]) 
\label{eq:genZ}
\\
&&\beta H_{eff}=
\frac{\lambda}{2} \int d^3{\bf r}\bigg(
\sum_{\mu \nu} E_{\mu\nu}^2
+\delta
\sum_{\nu} \bigg[ \sum_{\mu} \partial_{\mu}  E_{\mu\nu} \bigg]^2
+\eta \bigg[
\frac{1}{3} {\rm Tr}[{\bf E}]^2+\frac{1}{2} (E_{xy}-E_{yx})^2
+\frac{1}{2} (E_{xz}-E_{zx})^2
+\frac{1}{2} (E_{yz}-E_{zy})^2
\bigg]
\bigg)\nonumber\\
\label{eq:genHeff}
\end{eqnarray}
where the integral in Eq. (\ref{eq:genZ}) is taken independently over all components of the matrix
${\bf E}$ and $\beta=1/T$ is the inverse temperature.

The limit $\delta \to \infty, \eta\to0$, captures the Heisenberg model spin liquid, with Gauss'
law enforced on every column of ${\bf E}$ and no correlations between columns.
The  limit $\delta \to \infty, \eta\to\infty$ captures the R2-U(1) spin liquid, with ${\bf E}$ forced
to be symmetric and traceless and still obeying Gauss' law for each column.

In terms of the parameters of the microscopic model:
\begin{eqnarray}
\delta \sim \beta J, \quad  \eta \sim \beta |D_A|.
\end{eqnarray}
since $\delta$ is the coefficient enforcing the Gauss' law (which is generated by $J$) and
$\eta$ is the coefficient enforcing the symmetric and traceless conditions (which are generated by $D_A$).

We can study the crossover from the Heisenberg to R2-U(1) spin liquids by first taking $\delta\to \infty$
and observing the behavior as a function of $\eta$.
The crossover can be illustrated by calculating the correlation function
$$
\langle E^{yx}(\mathbf{q})  E^{yx}(-\mathbf{q}) \rangle
$$
which should have a 2-fold pinch point in the Heisenberg limit and a 4-fold pinch point in the R2-U(1)
limit.

Calculating the correlation function from Eqs. (\ref{eq:genZ})-(\ref{eq:genHeff})
and taking the limit $\delta \to \infty$.
gives us
\begin{eqnarray}
&&\langle E^{yx}(\mathbf{q})  E^{yx}(-\mathbf{q}) \rangle=
\frac{f(\mathbf{q}, \eta)}{2 \lambda q^4 (1+\eta)(2+\eta)(3+2\eta)} 
\label{eq:corrfuncyx}
\\
&&f(\mathbf{q}, \eta)=4 q_x^4 (1+\eta)(3+2\eta)+q_z^2(q_y^2+qz^2)(2+\eta)^2(3+2\eta)
+2 q_x^2 q_y^2 (1+\eta)(2+\eta)(3+\eta)
+q_x^2 q_z^2 (3+2\eta)(8+8\eta+\eta^2) .
\nonumber \\
\end{eqnarray}

The $\eta\to0$ limit of Eq. (\ref{eq:corrfuncyx}) gives the Heisenberg limit
of the correlation function (a 2-fold pinch point)
\begin{eqnarray}
\lim_{\eta \to 0}\langle E^{yx}(\mathbf{q})  E^{yx}(-\mathbf{q}) \rangle
=
\frac{1}{\lambda} \left( 1- \frac{q_y^2}{q^2} \right).
\end{eqnarray}

Whereas, the $\eta\to\infty$ limit gives the R2-U(1) correlation function
(a 4-fold pinch point):
\begin{eqnarray}
\lim_{\eta \to \infty}\langle E^{yx}(\mathbf{q})  E^{yx}(-\mathbf{q}) \rangle
=
\frac{1}{2\lambda} \frac{(q^2-q_x^2)(q^2-q_y^2)}{q^4}.
\end{eqnarray}

Since $\eta \sim \beta |D_A|$, we can see from Eq. (\ref{eq:corrfuncyx}) that
for a fixed ${\bf q}$  at any, fixed, finite temperature 
the evolution as a function of $D_A$, will be smooth, although at small
temperatures, a small change in $D_A$ will give a large change in $\eta$
and hence the correlation function.
Only at $T=0$ does the correlation function behave in a singular fashion as
a function of $D_A$, but this is not surprising, since the Heisenberg limit has
a highly degenerate ground state.

\begin{figure}
	\centering
	\includegraphics[width=0.5\columnwidth]{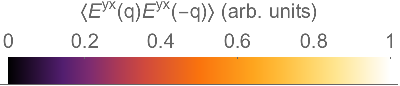}\\
	\includegraphics[width=0.22\columnwidth]{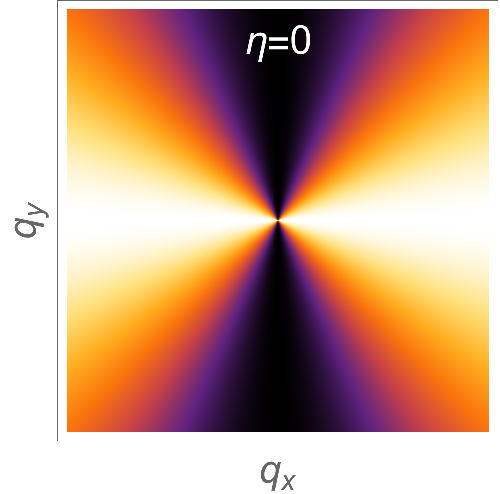}
	\includegraphics[width=0.22\columnwidth]{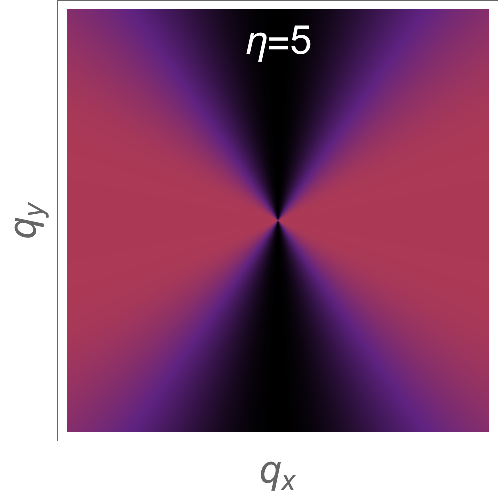}
	\includegraphics[width=0.22\columnwidth]{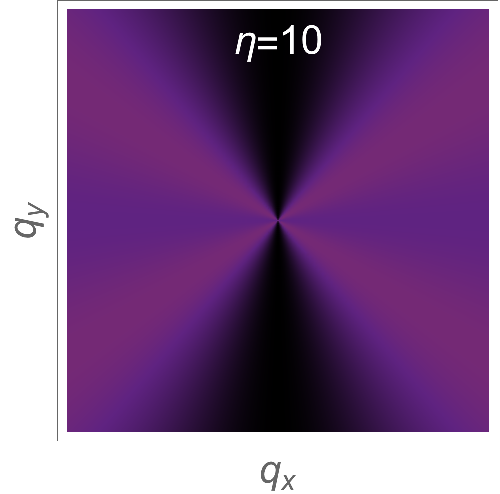}
	\includegraphics[width=0.22\columnwidth]{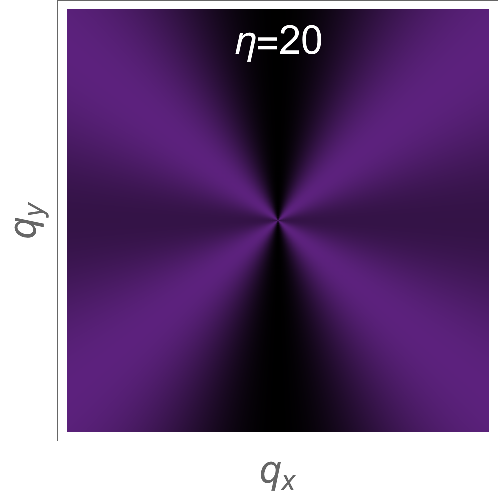}
	\caption{Evolution of the correlation function 
		$\langle E^{yx}(\mathbf{q})  E^{yx}(-\mathbf{q}) \rangle$
		[Eq. (\ref{eq:corrfuncyx})] 
		from a 2-fold pinch point to a 4-fold pinch point
		as we tune from the Heisenberg to R2-U(1) spin liquids by increasing
		the parameter $\eta$ [Eq. (\ref{eq:genHeff})].
		In terms of our microscopic model, this may be viewed either as 
		decreasing the temperature at fixed, small, $|D_A|$ or as increasing 
		$|D_A|$ at fixed small temperature.
	}
	\label{fig:eta_increasing}
\end{figure}

The progression of the correlation function as $\eta$ is increased is given
in Fig (\ref{fig:eta_increasing}). This progression can either be seen as 
decreasing the temperature at fixed, small, $|D_A|$ or as increasing 
$|D_A|$ at fixed small temperature.
\\

\section{Monte Carlo simulations}
\label{app:MC}

Monte Carlo simulations are performed on systems of classical O(3) spins with $16L^{3}$ sites, where $L^{3}$ is the number of cubic unit cells. The spin length is $|S|=1/2$. To decorrelate the system, we use jointly the heatbath method, over-relaxation and parallel tempering. Thermalization is made in two steps: first a slow annealing from high temperature to the temperature of measurement $T$ during $t_{e}$ Monte Carlo steps (MCs) followed by $t_{e}$ MCS at temperature $T$. After thermalization, measurements are done every 10 MCs during $t_{m}=10~t_{e}$ MCs. All structure factors have been computed from simulations with $L=30$ and $t_{m}=5\times 10^{5}$ MCs. The phase diagram of Fig. 5 has been computed from simulations with $L=8$ and $t_{m}=10^{7}$ MCs.






\end{document}